\documentclass[twoside,a4paper,11pt]{sca}
\usepackage{graphicx}
\usepackage{hyperref}
\usepackage{movie15}
\usepackage{natbib}  
\usepackage{floatflt}
\begin{document}
\pagenumbering{arabic}
\pagestyle{myheadings}
\thispagestyle{empty}
{\flushright\includegraphics[width=\textwidth,bb=90 650 520 700]{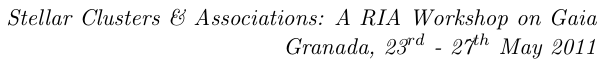}}
\vspace*{0.2cm}
\begin{flushleft}
{\bf {\LARGE
%
Observing brown dwarfs in the Magellanic Cloud star-forming regions with the E-ELT
%
}\\
\vspace*{1cm}
%
Annalisa Calamida$^{1}$,
Fernando Comeron$^{2}$, 
and 
Hans Zinnecker$^{3}$
%
}\\
\vspace*{0.5cm}
%
$^{1}$
Osservatorio Astronomico di Roma (INAF)\\
$^{2}$ESO\\
$^{3}$ìNASA Ames
\end{flushleft}
%
\markboth{
Observing brown dwarfs in the Magellanic Clouds with the E-ELT
}{ 
Calamida et al.
}
\thispagestyle{empty}
\vspace*{0.4cm}
\begin{minipage}[l]{0.09\textwidth}
\ 
\end{minipage}
\begin{minipage}[r]{0.9\textwidth}
\vspace{0.8cm}
\section*{Abstract}{\small
We present the results of near-infrared imaging simulations
of young star-forming regions in the Magellanic Clouds
to be observed with the European Extremely Large Telescope (E-ELT). 
The simulated $J,H,K$-band images show that we should be able to 
obtain nearly complete samples of young brown dwarfs  
above the deuterium burning limit ($M >$ 13 $M_{Jup}$)
in low-mass star-forming regions in the Clouds.
Moreover, very young giant planet-mass objects in the Clouds
should be detectable with the E-ELT under favourable conditions. 
\normalsize}
\end{minipage}
%
\vspace{-0.35cm}
\section{Magellanic Cloud brown dwarfs in the E-ELT Design Reference Mission\label{intro}}
\vspace{-0.35cm}
One of the projects proposed as part of the E-ELT Design Reference Mission (DRM)
is the determination of the low-mass luminosity function down to the giant 
planet-mass regime in low-metallicity star-forming regions of the Large 
Magellanic Cloud (LMC, $[Fe/H] = -0.33$ dex) and the Small Magellanic 
Cloud (SMC, $[Fe/H] = -0.75$ dex, \citealt{roma}). 
Typical low-mass star-forming regions of the solar neighborhood, such as 
Lupus 3 or $\rho$-Ophiuchi, would subtend an angle of $\sim$ 2" at the distance 
of the Clouds, thus being appropriate for Laser Tomography Adaptive Optic 
(LTAO) observations, which would provide nearly diffraction-limited, 
deep near-infrared (NIR) imaging of the whole stellar and sub-stellar content 
of the cluster. 
In order to explore the detectability of the lowest-mass objects we 
simulate very  young clusters similar to those found in the solar vicinity, 
without massive stars ($M <$ 2 $M_{\odot}$) able to significantly ionize the gas, 
as this would dramatically increase the NIR background brightness via 
free-free emission.
The challenge of the observations is the high density of 
objects ($\sim$ 20 objects/arcsec$^2$) and the fact that the members of 
the star-forming region cover a wide range in magnitudes, from $K \sim 18$ to $K \sim$ 30 mag.

\begin{figure}[htbp]
     \begin{minipage}[l]{0.48\textwidth}
      \centering
\hspace{-0.5truecm}
       \includegraphics[height=6.cm,width=6.7cm]{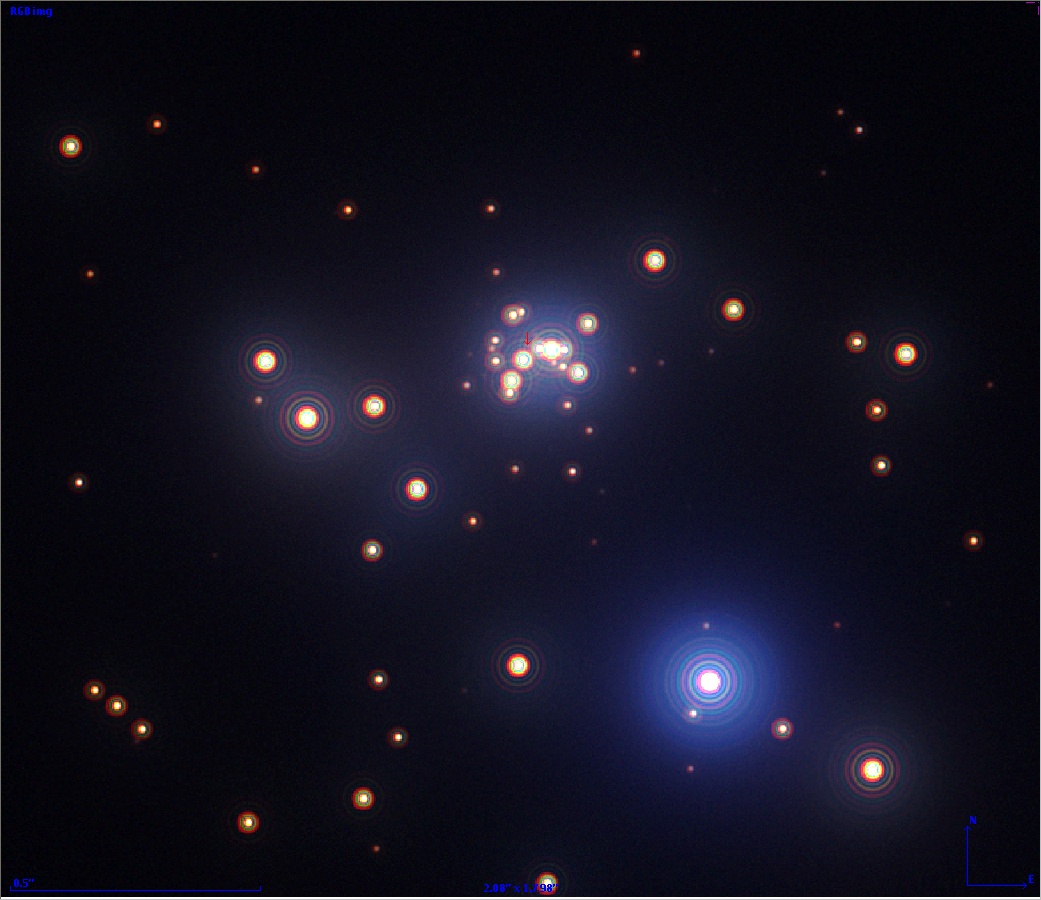}
\vspace{-0.truecm}
       \caption{\label{fig1} Composite $J,H,K$-band image of a 
       low-mass star-forming region in the LMC, as observed with E-ELT and
       LTAO. 100 objects are simulated with masses in the range 
       5$M_{Jup} < M < 2M_{\odot}$, randomly distributed in a field 
       of view of 2"$\times$2" (0.5 pc $\times$ 0.5 pc).}
     \end{minipage}\hfill
     \begin{minipage}[r]{0.48\textwidth}
      \centering
\hspace{-1.2truecm}
\vspace{1.6truecm}
       \includegraphics[height=6.5cm,width=7cm]{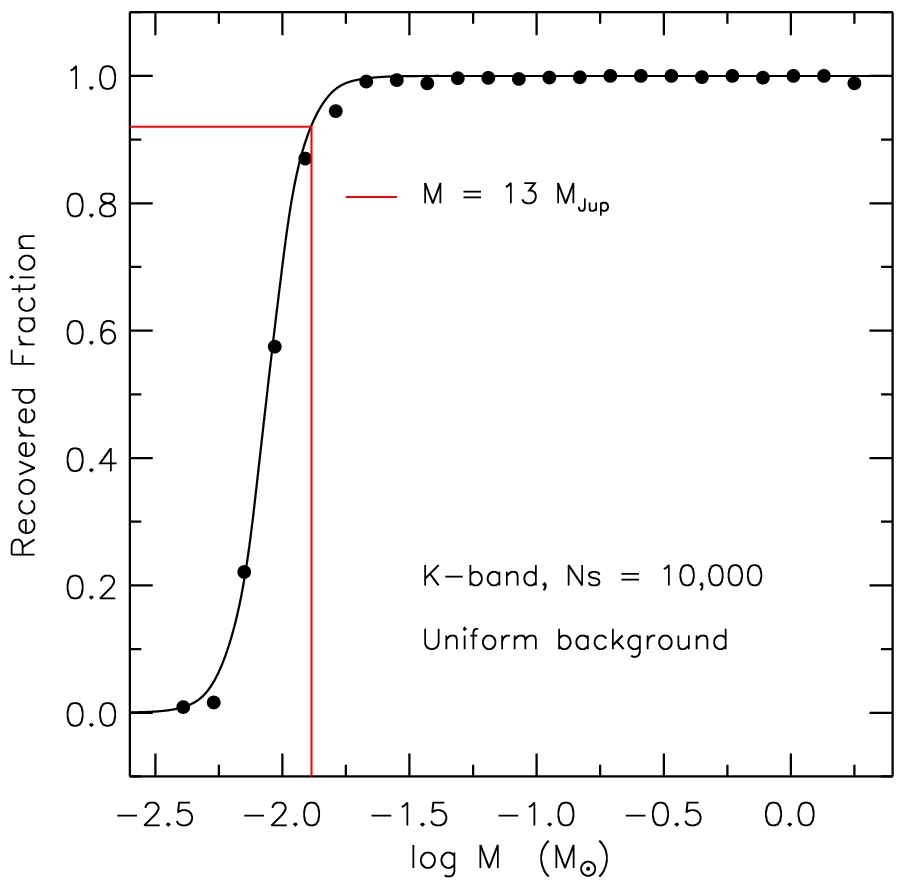}
\vspace{-2truecm}
       \caption{\label{fig2} Recovered fraction of stars versus the input mass 
       in steps of 0.12 $M_{\odot}$.  
       The completeness function is overplotted as a solid line. 
       The red line indicates the deuterium burning limit, $M$ = 13 $M_{Jup}$, 
       where the recovered fraction of objects is $\sim$ 90\%.}
     \end{minipage}
   \end{figure}

We simulate $J,H,K$-band images assuming different zenith distance values, 
in order to assess how much the location of the telescope will influence 
the observations, and different instrument pixel scales and stellar densities. 
We assume a distribution of stellar masses drawn from \citet{chabrier2003} 
initial mass function truncated at 2 $M_{\odot}$.
Stars are randomly distributed over the 2"$\times$2" extent of the cluster 
and masses are transformed to $J,H,K$ absolute magnitudes using a 
$t = 5$ Myr isochrone from \citet{baraffe2003}.
A distance modulus of $18.5$ mag is added plus a random 
extinction for each star. Variable background is introduced in some cases. All the 
simulations are performed adopting the technical assumptions of the official 
DRM data base (http://www.eso.org/sci/facilities/eelt/science/drm/tech\_data).
Fig.~\ref{fig1} shows a composite $J,H,K$ simulated image of a young star-forming
region in the LMC.

We then perform crowded field photometry on the simulated images and compare 
the recovered luminosity function with the input one, in order to verify 
if giant planet-mass objects can be detected and measured with sufficient 
accuracy, i.e. $S/N >$ 5 (see Fig.~\ref{fig2}).

\vspace{-0.5cm}
\section{Conclusions \label{concl}}
\vspace{-0.2cm}
Our simulations show that nearly complete samples of young brown dwarfs 
above the deuterium burning limit ($M >$ 13 $M_{Jup}$) can be obtained in the 
Magellanic Cloud low-mass star-forming regions with E-ELT, and 
even giant planet-mass objects may be detected if the cluster nebulosity 
is sufficiently faint.

\small  
%
%

%
%
%
%
%

\bibliographystyle{aa}
\bibliography{mnemonic,ref_calamida_1}

\end{document}